\documentclass[a4paper]{article}
\usepackage[doublespacing]{setspace}
\usepackage{amsbsy,amssymb,amsmath}
\usepackage{graphicx}
\usepackage{eucal}
\newcommand{\hana}[1]{{\mathcal{#1}}}
\newcommand{\dd}{{\rm d}}

\newcommand{\eexp}{{\rm e}}
\newcommand{\Tr}{{\begin{subarray}{c}\mbox{\large Tr}\\\{x_i^a\}\end{subarray}}
}
\newcommand{\Trna}{{\begin{subarray}{c}\mbox{\large Tr}\\\{x_i\}\end{subarray}}
}
\newcommand{\Trni}{{\begin{subarray}{c}\mbox{\large Tr}\\\{x^a\}\end{subarray}}
}
\def\bm#1{\mbox{\boldmath $#1$}}

\title{Statistical mechanics of relative species abundance}

\author{Kei Tokita\\
Large-Scale Computational Science Division, 
Cybermedia Center, Osaka University\\
Graduate School of Science, Osaka University\\
Graduate School of Frontier Biosciences, Osaka University\\
1-32 Machikaneyama-cho, Toyonaka, Osaka 560-0043, Japan\\
(tokita@cmc.osaka-u.ac.jp; http://www.cp.cmc.osaka-u.ac.jp/\%7Etokita/)\\
}

\date{\today}

\begin{document}
\doublespacing
\maketitle

\begin{abstract}
Statistical mechanics of relative species abundance (RSA)
patterns in biological networks is presented. The theory is based on
multispecies replicator dynamics equivalent to the Lotka-Volterra
equation, with diverse interspecies interactions. Various RSA patterns
observed in nature are derived from a single parameter related to
productivity or maturity of a community. The abundance distribution is formed like a
widely observed left-skewed lognormal distribution. It is also found
that the ``canonical hypothesis'' is supported in some parameter region
where the typical RSA patterns are observed.  As the model has a general form,
the result can be applied to similar patterns in other complex
biological networks, e.g. gene expression.
\end{abstract}

\newpage

%
%
\section{Macroscopic ecological patterns as eco-information}
The most significant feature of large-scale biological networks, such as
food webs\cite{Martinez_1991,LittleRockLake}Cmetabolic networks in a
cell\cite{BCPathway} and protein
networks\cite{Jeong_et_al_2001,Gallery}, is overwhelming diversity of
components, e.g., species, chemical constituents and proteins,
respectively, great complexity of network topology, and homeostatic
stability of dynamics on the networks. From a theoretical viewpoint, it
is a serious question how living organisms has evolved such a
homeostasis because chaotic instability is inherited even in a simple
nonlinear system.

As an approach to such a problem, macroscopic patterns observed in
various complex networks have been
studied\cite{Barabasi_Linked_2002}. Such studies on {\it
scale-free networks} have been elucidated the characteristics of
topology of natural and artificial complex networks, and the
evolutionary conditions which produces such a topology. Unifying
approaches to biological and abiological networks have emphasized their
similarity and difference. For example, it is pointed out that an
infection of computer virus on the internet with scale-free topology is
followed by a qualitatively different epidemic dynamics from the one of
biological viruses in nature, and, therefore, the computer viruses are
hardly eradicated\cite{Pastor-Satorras_Vespignani_2001}.

On the other hand, in a large-scale complex biological networks such as
ecosystems, not only a topology of the network links but also a
thickness of each link, i.e. the strength of interactions, definitely
affect population dynamics and resulting macroscopic patterns. In
ecology, classical macroscopic patterns observed and studied for a long
time is RSA patterns, in other words, abundance distribution
 of species, which is one of the most
accumulated informations obtained in ecology.

Nevertheless, how to clarify the mechanisms underlying those RSA patterns
has been one of the 'unanswered questions in ecology in the last
century \cite{May_1999}' even though the knowledge obtained from it
would affect vast areas of nature conservation. Various models have been
applied to ecosystem communities where species compete for niches on a
trophic level
\cite{Motomura_1932,Corbet_Fisher_Williams_1943,MacArthur_1957,MacArthur_1960,Preston_1962a,Preston_1962b,Whittaker_1970,Bazzaz_1975,May_1975,Sugihara_1980,Nee_Harvey_May_1991,Tokeshi_1999,Hubbell_2001,Hall_etal_2002,Harte_2003,McGill_2003,Volkov_etal_2003,Pigolotti_etal_2004,Etienne_Olff_2004,Chave_2004,Harte_etal_EcolMonog_2005},
but these models have left the more complex systems a mystery. Such
systems occur on multiple trophic levels and include various types of
interspecies interactions, such as prey-predator relationships,
mutualism, competition, and detritus food chains. Although RSA patterns are
observed universally in nature, their essential parameters have not been
fully clarified. In this paper, it is presented that 
RSA patterns are derived from a statistical mechanical
theory\cite{Tokita_2004}, based on a general evolutionary dynamics which
is applied in vast area of fields.

We consider here the {\it replicator
equation}\cite{Hofbauer_Sigmund_1998} (RE),
\begin{eqnarray}
&&\frac{{\rm d}x_i}{{\rm d}t}=x_i \left(f_i(\bm{x})-\frac{1}{N}\bar{f}(\bm{x})\right)_,\label{eqReplicator}\\
&&f_i(\bm{x})\equiv\sum_j^N J_{ij}x_j,\nonumber\\
&&\bar{f}(\bm{x})\equiv\sum_i^Nf_i(\bm{x}) x_i\nonumber
\end{eqnarray}
where $N$ is the number of species, and $0\leq x_i(t)\leq N$ denotes $i$th
species' population. The functions $f_i(\bm{x})$ and $\bar{f}(\bm{x})$ 
denote {\it fitness} of species $i$ and it's average, respectively.
Interaction between $i$th species and $j$ is specified by $J_{ij}$.
Note that total population is conserved at any time as
$\sum_i x_i=N$ and, that is, the trajectory of the dynamics 
(\ref{eqReplicator}) is bounded in a simplex $\sum_i x_i=N$.

The RE appears in various fields \cite{Hofbauer_Sigmund_1998}. In
sociobiology, it is a game dynamical equation for the evolution of
behavioral phenotypes; in macromolecular evolution, it is the basis of
autocatalytic reaction networks ({\it hypercycles}); and in population
genetics it is the continuous-time selection equation in the symmetric
$(J_{ij}=J_{ji} )$ case. 
The symmetric RE also corresponds to a classical
model of competitive community for
resources\cite{MacArthur_Levins_1967}.
The replicator dynamics, therefore, are often used as a model of
complex systems in which many components changes their numbers
through complex reaction, replication and reproduction of the
components.

Here we assume that $(J_{ij})$ is a
time-independent random symmetric $(J_{ij}=J_{ji})$ matrix whose
elements have a normal distribution with mean $m (>0)$ and variance
$\tilde{J}^2/N$ as
\[
P(J_{ij})=\sqrt{\frac{N}{2\pi\tilde{J}^2}}\exp\left[
               - \left(
                   \frac{N}{2\tilde{J}^2}
                \right)\left(J_{ij}-m\right)^2
                                              \right]\qquad (i\neq j)
\]
Self-interactions are all set to a negative constant as $J_{ii}=-u
(<0)$.  Note that the essential parameter is unique as $p\equiv
(u+m)/\tilde{J}$ because the transformation of the interaction
$K_{ij}\equiv (J_{ij}-m)/\tilde{J}$ does not change the trajectory of
the dynamics (\ref{eqReplicator}).  Although ecologists do not generally
believe in the randomness of interspecies interactions in nature, the
discipline has been affected by the random interaction model
\cite{May_1972} as a prototype of complex systems.

 Particularly in the context of ecology, the $N$
 species RE \cite{Hofbauer_Sigmund_1998} is equivalent to the
$N-1$ species Lotka-Volterra (LV) equation
\[
 \frac{\dd y_i}{\dd t}=y_i\left(
                       r_i - \sum_j^{N-1}b_{ij}y_j
                          \right)_.\label{eqLV}
\]
That is, the abundance $y_i$ and the parameters
in the corresponding LV are described by those in
the present RE model as,
\begin{eqnarray}
&&y_i = x_i/x_M \quad (i=1,2,\ldots ,N),\label{yi}\\
&&r_i = J_{iM} - J_{MM} = J_{iM} + u,\label{ri}\\
&&b_{ij} = J_{ij}-J_{Mj}\label{bij}
\end{eqnarray}
where the 'resource' species $M\, (y_M=1)$ can be arbitrarily chosen
from $N$ species in the RE. The ecological interspecies interactions
$(b_{ij})(i\neq j)$ have a normal distribution with mean $0$ and
variance $2\tilde{J}^2/N$ from Eq.~(\ref{bij}), and they are no longer
symmetric ($b_{ij}\neq b_{ji}$). The present model therefore describes
an ecological community with complex prey-predator interactions
$((b_{ij}, b_{ji})\to (+,-)\,\mbox{or}\, (-,+))$, mutualism $(+,+)$ and
competition $(-,-)$. Moreover, a community can have a 'loop' (detritus)
food chain $((b_{ij}, b_{ji})\to (+,-), (b_{jk}, b_{kj})\to (+,-),
(b_{ki}, b_{ik})\to (+,-))$.  The intraspecific interaction $b_{ii}$
turns out to be related to the intrinsic growth rate $r_i$ as
$b_{ii}=J_{ii}-J_{Mi}=-u-J_{iM}=-r_i$ and is therefore competitive
$(b_{ii}<0)$ for producers $(r_i>0)$ or mutualistic $(b_{ii}>0)$ for
consumers $(r_i<0)$.

By Eq.~(\ref{ri}), the intrinsic growth rates also have a normal
distribution with mean $u+m$ and variance $\tilde{J}^2/N$. The
probability at which $r_i$ is positive--that is, that the
$i$-th species is a producer--is therefore given by the
error function,
\[
 \mbox{Prob}(r_i>0)=\int_{-p\sqrt{N/2}}^\infty 
   \frac{\dd t}{\sqrt{\pi}}\exp\left(-t^2\right)_.
\]
Consequently, the parameter $p$ can be termed as the 'productivity' of
a community because the larger the $p$, the greater the number of
producers. This can be also understood from the fact that $p$ is connected
to the average growth rate:
\begin{equation}
 \frac{1}{N}\sum_ir_i=\left\langle J_{iM}+u\right\rangle_J=m+u=p\tilde{J}.\label{defp}
\end{equation}
The parameter $p$ is also connected to the maturity of an
ecosystem because $m$ increases in time in an evolutionary model
\cite{Tokita_Yasutomi_2003}.

Note that the growth rate $(1/x_i)\dd x_i/\dd t$ in RE
(\ref{eqReplicator}) has no ecological meaning because it is defined by
the average fitness $\bar{f}$ subtracted from the fitness $f_i$. We
therefore consider the equivalent LV (\ref{eqLV}) when we discuss the
model in the context of ecology as stated above. Why we do not
consider LV with random asymmetric interactions from the beginning is
that the system we consider here is not general asymmetric LVs but a
class of LVs which is corresponded to a symmetric RE whose symmetry is
crucial to the present analysis.

While random asymmetric interaction matrix ($J_{ij}\neq
J_{ji}\,\mbox{and},\, J_{ij}\, \mbox{and}\, J_{ji}\, \mbox{are independent each other}$) was
assumed in the classical random population
models\cite{Gardner_Ashby_1970,May_1972,Tokita_Yasutomi_1999}, here, the
symmetric matrix ($J_{ij}=J_{ji}$) enables us to derive 
RSA patterns and {\it left-skewed}\cite{Nee_Harvey_May_1991}
{\it canonical}\cite{Preston_1962b,May_1975} lognormal-like distribution,
from a single parameter $p$\cite{Tokita_2004}.

%
%
\section{Mean field theory of random symmetric replicator dynamics}

The symmetry $(J_{ij}=J_{ji})$ makes the average fitness
$\bar{f}\equiv\sum_{j,k}^NJ_{jk}x_jx_k$ (the second term of the
r.h.s. of Eq.~(\ref{eqReplicator})) a Lyapunov function (Appendix A)
\cite{Hofbauer_Sigmund_1998}, which is a nondecreasing function of time
in dynamics (\ref{eqReplicator}). Therefore, every initial state converges to a
local maximum of $\bar{f}$ as $t\to\infty$. Interpreting $\hana{H}\equiv
-\frac{1}{2}\bar{f}$ as an energy function, macroscopic (thermodynamic)
functions of the system is derived from free energy 
\begin{equation}
f=-\lim_{\beta\to\infty}\lim_{N\to\infty}\frac{\left\langle\ln Z\right\rangle_J}{N\beta}\label{free_energy}
\end{equation}
at such a maximum by
using the technique of statistical mechanics of random systems
\cite{Mezard_etal_1987,Diederich_Opper_1989,Biscari_Parisi_1995,de_Oliveira_Fontanari_2000,de_Oliveira_Fontanari_2001,de_Oliveira_Fontanari_2002}.
The bracket 
\begin{equation}
\left\langle F(J_{ij})\right\rangle_J\equiv\left(\prod_{i<j}^N\int_{-\infty}^\infty \dd J_{ij}P(J_{ij})\right)F(J_{ij})\label{RandomAverage}
\end{equation}
denotes the {\it random
average}\cite{Mezard_etal_1987} over random interactions, by which a
typical behavior of the system can be analyzed.
The normalization factor $Z$ denotes a partition function with the condition
($\sum_i x_i=N$) and is represented as
\begin{equation}
Z\equiv  \int_0^\infty \prod_{i=1}^N\dd
 x_i \delta(N-\sum_ix_i)\eexp^{-\beta\hana{H}}
 \equiv \Trna\eexp^{-\beta\hana{H}},\label{Partitionfunc}
\end{equation}
where {ensemble average} is represented by the {\it trace} $\Trna$.

Why we execute the random average is that free energy 
is ``self-averaging'', that is, it is represented by
an average over random interactions, not by a detail of each sample
of interactions, which is justified in the thermodynamic limit,
where the number of species are very large in the context of ecology.
The inequality $i<j$ in (\ref{RandomAverage}) denotes the product of the
values of  $i$ and $j$ satisfying $1\leq\cdots <i<j<\cdots\leq N$.

If we define Hamiltonian as
\begin{equation}
\hana{H}\equiv -\frac{1}{2}\sum_{i,j}
 J_{ij}x_i x_j+h\sum_i\theta (y-x_i),\label{Hamiltonian}
\end{equation}
where $\theta(z)[=1(z\geq 0); 0(z<0)]$ is the step function, 
we will be able derive cumulative distribution function $C(y)$
(proportion of the number of species which has abundance less than and
equal to $y$) as
\begin{eqnarray}
C(y)
&\equiv &\lim_{\beta\to\infty}\lim_{N\to\infty}\lim_{h\to 0}\left\langle
 \frac{1}{Z}\Trna\frac{\sum_i\theta(y-x_i)}{N}\eexp^{-\beta\hana{H}}
\right\rangle_J
=-\lim_{\beta\to\infty}\lim_{N\to\infty}\lim_{h\to 0}
   \frac{1}{N\beta}\frac{\partial\langle\ln Z\rangle_J}{\partial h}\nonumber\\
&=&\lim_{h\to 0}\frac{\partial f}{\partial h},\label{Cy}
\end{eqnarray}
where $\theta (x)[=1(x>0); 0(x\leq 0)]$ is the step function.
Information of equilibrium of RE (\ref{eqReplicator}) can be obtained by setting $h=0$.
By the identical equation
\begin{equation}
\langle\ln Z\rangle_J\equiv \lim_{n\to 0}\frac{\langle Z^n\rangle_J
 -1}{n}\label{replica}
\end{equation}
the random average of the logarithm of the partition function, which is
hard to execute analytically, can be transformed to the random average
of an equivalent $n$-replicated partition function
\begin{equation}
\left.\langle
Z^n\rangle_J=\left\langle\prod_{a=1}^n\Tr\eexp^{-\beta\hana{H}_a}\right\rangle_J
=
\left\langle\prod_{a=1}^n\Tr
\exp\left\{-\beta\left(
		  -\frac{1}{2}\sum_{i,j}
		  J_{ij} x_i^a x_j^a+h\sum_i\theta (y-x_i^a)
                 \right) 
    \right\}
\right\rangle_J\right._,\nonumber
\end{equation}
which is more tractable. Each $\hana{H}_a$ denotes a {\it replica
Hamiltonian} where an variable $x_i$ is replaced by $x_i^a$ with a
replica index $a=1, 2, \dots, n$ in Eq. (\ref{Hamiltonian}). The
analysis using above transformation of the Hamiltonian is called as the
{\it replica method}\cite{Mezard_etal_1987} which enables us to analyze
typical behavior of free energy with random interactions.  The replica
method was originally invented for analysis of the {\it spin glass},
magnetic alloy. Recently it has been successfully applied to various
models\cite{Nishimori_2001} with time-invariant random interactions
other than physical systems. By exchanging the order of the trace and
the random average, we can precedently execute the random average.  As
the integrals (\ref{RandomAverage}) are Gaussian integral of $N(N-1)/2$
variables $J_{ij}$, the replicated partition function becomes
\begin{equation}
\langle Z^n\rangle_J=
\Tr\exp\beta\left[
 \frac{\beta \tilde{J}^2}{2N}\sum_{i<j}\left(\sum_ax_i^ax_j^a\right)^2 +
 \sum_a\sum_{i<j}mx_i^ax_j^a
 -\frac{u}{2}\sum_a\sum_i(x_i^a)^2 
     - h\sum_a\sum_i\theta(y-x_i^a)
\right]_.\nonumber
\end{equation}
As the first term in $[\cdots]$ above can be rewritten as following,
\begin{equation}
\sum_{a<b}\left(\sum_ix_i^ax_i^b\right)^2 
- \frac{1}{2}\sum_i\left[\sum_a(x_i^a)^2\right]^2
+ \frac{1}{2}\left[\sum_i(x_i^a)^2\right]^2_,\nonumber
\end{equation}
we can derive
\begin{eqnarray}
\langle Z^n\rangle_J&=&\Tr\exp\beta\Bigg\{
 \underbrace{
  \frac{\beta \tilde{J}^2}{2N}\sum_{a<b}\left(\sum_ix_i^ax_i^b\right)^2
 }_{K_1}
 -\frac{\beta \tilde{J}^2}{4N}
   \sum_i\left[\sum_a(x_i^a)^2\right]^2\nonumber\\
& &\mbox{} + \underbrace{
              \frac{\beta \tilde{J}^2}{4N}
              \sum_a\left[\sum_i(x_i^a)^2\right]^2
             }_{K_2}
+ \underbrace{
    m \sum_a\sum_{i<j}x_i^ax_j^a
  }_{K_3}
- \frac{u}{2}\sum_a\sum_i(x_i^a)^2
 - h \sum_a\sum_i\theta(y-x_i^a)\Bigg\}_.\nonumber
\end{eqnarray}
As we rewrite like following,
\begin{equation}
K_3=\frac{m}{2}\sum_a\Bigg\{
     \underbrace{
      \left(\sum_ix_i^a\right)^2
     }_{K_4}
     - \sum_i\left(x_i^a\right)^2
                     \Bigg\}\nonumber
\end{equation}
and apply the Hubbard-Stratonovich transformation
\begin{equation}
\eexp^{\lambda a^2}\equiv
\frac{1}{\sqrt{2\pi}}\int_{-\infty}^\infty
 \dd x\exp\left(-\frac{x^2}{2}+\sqrt{2\lambda}ax\right)\nonumber
\end{equation}
where $\lambda >0$, the quadratic terms of 
$(\sum_i\cdots)$ in $K_1, K_2, K_4$ can be transformed to 
linear terms. By this, we can execute the trace $\Tr$ and obtain
\begin{eqnarray}
&&\!\!\!\!\!\!\!\!\!\!\!\!\!\!\!\exp(\beta K_1)=\prod_{a<b}\exp\frac{(\beta
 \tilde{J})^2}{2N}\left(\sum_ix_i^ax_i^b\right)^2
=\prod_{a<b}\frac{1}{\sqrt{2\pi}}
  \int_{-\infty}^{\infty}dy_{ab} \exp\left\{
     -\frac{y_{ab}^2}{2}+\frac{\beta \tilde{J}}{\sqrt{N}} \sum_ix_i^ax_i^by_{ab}
                         \right\}\nonumber\\
&&\!\!\!\!\!\!\!\!\!\!\!\!\!\!\!\exp(\beta K_2)=\prod_a\exp\frac{(\beta
 \tilde{J})^2}{4N}\left[\sum_i(x_i^a)^2\right]^2
=\prod_a\frac{1}{\sqrt{2\pi}}
  \int_{-\infty}^{\infty}ds_a \exp\left\{
     -\frac{s_a^2}{2}+\frac{\beta \tilde{J}}{\sqrt{2N}} \sum_i(x_i^a)^2s_a
                         \right\}\nonumber\\
&&\!\!\!\!\!\!\!\!\!\!\!\!\!\!\!\exp(\beta K_4)=\prod_a\exp
   \frac{m\beta}{2}\left(\sum_ix_i^a\right)^2
=\prod_a\frac{1}{\sqrt{2\pi}}
  \int_{-\infty}^{\infty}dt_a \exp\left\{
     -\frac{t_a^2}{2}+\sqrt{m\beta} \sum_ix_i^at_a
                         \right\}_.\nonumber
\end{eqnarray}
If we transform the variables as $y_{ab}\equiv\beta\sqrt{N}Y_{ab}, s_a\equiv\beta\sqrt{N}S_a, t_a\equiv\beta\sqrt{N}T_a$, we obtain
\begin{eqnarray}
\!\!\!\!\!\!\!\!\!\!\!\!\!\!\!\!\!\!\!\!\!\!\!\!\!\!\!\!\!\!\!\!
\left\langle Z^n\right\rangle_J 
 &=&
 \Tr\left[
     \prod_{a<b}\int_{-\infty}^\infty\frac{\dd Y_{ab}}{L}
    \right]
    \left[
     \prod_a\int_{-\infty}^\infty\frac{\dd  S_a}{L}
    \right]
    \left[
     \prod_a\int_{-\infty}^\infty\frac{\dd T_a}{L}
    \right]\nonumber\\
&\times&
 \exp\beta\Bigg[
      -\left(\frac{N\beta}{2}\right)\sum_{a<b}Y_{ab}^2
      +\beta \tilde{J}\sum_{a<b}Y_{ab}\sum_ix_i^ax_i^b\nonumber\\
&&\mbox{}\qquad\quad
      -\left(\frac{N\beta}{2}\right)\sum_aS_a^2
      +\left(\frac{\beta
	\tilde{J}}{\sqrt{2}}\right)\sum_aS_a\sum_i(x_i^a)^2\nonumber\\
&&\mbox{}\qquad\quad
      -\left(\frac{N\beta}{2}\right)\sum_aT_a^2
      +\sqrt{mN\beta}\sum_aT_a\sum_ix_i^a\nonumber\\
&&\mbox{}\qquad\quad
      -\left(\frac{u+m}{2}\right)\sum_a\sum_i(x_i^a)^2
      -h\sum_a\sum_i\theta(y-x_i^a)
     \Bigg]_,\nonumber
\end{eqnarray}
where $L\equiv\sqrt{2\pi/\beta^2N}$. Let us write $g_{ab}$
for the terms with $\sum_i$ in $[\cdots]$. 
The delta function in the trace (\ref{Partitionfunc}) can be represented as a Fourier 
transformation 
\begin{equation}
\Tr
\Rightarrow\prod_{i,a}\int_0^\infty\dd x_i^a
             \int_{-i\infty}^{i\infty}\frac{\dd r_a}{2\pi i}
              \exp\left\{-\sum_ar_a(\sum_ix_i^a-N)\right\}.\nonumber
\end{equation}
The terms including $g_{ab}$ and the terms of $\sum_i$ above
can be represented by a product of independent terms, and therefore
can be written by the $N$th power of a term in which index $i$ is
omitted
as
\begin{equation}
\Tr\exp\left(g_{ab}-\sum_ar_a\sum_ix_i^a\right)
   =
    \Bigg[
    \underbrace{(\prod_a\int_0^\infty\dd
     x^a)\exp\left(g'_{ab}-\sum_ar_ax^a\right)
    }_A\Bigg]^N
=\exp\ln[A]^N=\exp N\ln[A].\nonumber
\end{equation}
The term $g'_{ab}$ denotes $g_{ab}$ without index $i$.
Now we come to sum up the $n$-replicated partition function
averaged over samples as
\begin{equation}
\left.\left\langle Z^n\right\rangle_J 
 =
 \left[
     \prod_{a<b}\int_{-\infty}^\infty\frac{\dd Y_{ab}}{L}
    \right]
    \left[
     \prod_a\int_{-\infty}^\infty\frac{\dd S_a}{L}
    \right]
    \left[
     \prod_a\int_{-\infty}^\infty\frac{\dd T_a}{L}
    \right]
    \left[
     \prod_a\int_{-i\infty}^{i\infty}\frac{\dd r_a}{2\pi i}
    \right]
    \eexp^{NG\{Y,S,T,r\}}\right._,\label{Z1}
\end{equation}
where
\begin{equation}
G\{Y,S,T,r\}=-\frac{\beta^2}{2}\left(
               \sum_{a<b}Y_{ab}^2+\sum_aS_a^2+\sum_aT_a^2
                               \right)+\sum_ar_a
+\ln\left[
		 \Trni
		 \exp\left(g'_{ab}-\sum_ar_ax^a\right)
		\right]\nonumber
\end{equation}
and $\Trni\equiv\left(\prod_a\int_0^\infty\dd x^a\right)$.
By the saddle point method, the integral (\ref{Z1}) can be replaced
by the integrand $\exp(NG)$, and by substituting this for
(\ref{free_energy}) and (\ref{replica}), the free energy 
can be represented as
\begin{eqnarray}
f&=&-\lim_{N\to\infty}\lim_{n\to 0}\left(\frac{1}{\beta Nn}\right)
 \left\{\exp(-\beta N{\rm Min}f_n)-1\right\}\nonumber\\
 &=&
-\lim_{N\to\infty}\lim_{n\to 0}\left(\frac{1}{\beta Nn}\right)
 \left\{\exp\left(-\beta Nn\lim_{n\to 0}\frac{{\rm
	     Min}f_n}{n}\right)-1\right\}\nonumber\\
 &=&-\lim_{N\to\infty}\lim_{n\to 0}\left(\frac{1}{\beta Nn}\right)
 \left\{\left(1-\beta Nn\lim_{n\to 0}\frac{{\rm
	     Min}f_n}{n}\right)-1\right\}\nonumber\\
 &=&\lim_{n\to 0}\frac{1}{n}{\rm Min}f_n\nonumber
\end{eqnarray}
by the minimum of $f_n$. If we transform the variables as
$q_{ab}=Y_{ab}/\tilde{J},\, \sigma_a=S_a\sqrt{2}/\tilde{J},\, \tau_a=T_a\sqrt{\beta/Nm}$,
the free energy becomes
\begin{equation}
\beta f_n=\sum_{a<b}\frac{(\beta \tilde{J})^2}{2}q_{ab}^2
          +\sum_a\frac{(\beta \tilde{J})^2}{4}\sigma_a^2
          +\sum_a\frac{\beta Nm}{2}\tau_a^2
          - \sum_ar_a 
          -\ln\left[\Trni\eexp^{-\beta\hana{H}_n^{eff}(x^a)}\right]_,\label{fn1}
\end{equation}
where
\begin{eqnarray}
\hana{H}_n^{eff}(x^a)&\equiv &
 -\beta \tilde{J}^2\sum_{a<b}x^ax^bq_{ab}
 -\frac{\beta \tilde{J}^2}{2}\sum_a(x^a)^2\sigma_a
 +Nm\sum_ax^a\tau_a
 +\frac{1}{\beta}\sum_ar_ax^a\nonumber\\
& &\mbox{} 
 +\frac{u+m}{2}\sum_a(x^a)^2
 + h\sum_a\theta(y-x^a).\nonumber
\end{eqnarray}

Here we assume the {\it Replica Symmetry (RS)} as
\begin{equation}
q=q_{ab}, \quad \sigma=\sigma_a, \quad \tau=\tau_a, \quad r=r_a\quad 
\mbox{for}\quad \forall a,b.\nonumber
\end{equation}
By the discussion on the stability of saddle point solution for the
estimation of the free energy, RS is justified at least for
$p\ge\sqrt{2}$ \cite{Diederich_Opper_1989,Biscari_Parisi_1995}.
By the RS order parameters, we can write as
\begin{eqnarray}
g^{RS}\equiv -\beta\hana{H}_n^{eff}(x^a)
  &=&\underbrace{\frac{(\beta \tilde{J})^2}{2}q}_{B_1}\left(\sum_ax^a\right)^2
  \underbrace{-\beta\left(\frac{\beta \tilde{J}^2}{2}q-\frac{\beta
		     \tilde{J}^2}{2}\sigma+\frac{u+m}{2}\right)}_{B_2}\sum_a(x^a)^2
\nonumber\\
 &&
  -\underbrace{(r+N\beta m\tau)}_{B_3}\sum_ax^a
          -\underbrace{\beta h\sum_a\theta(y-x^a).}_{B_4}\nonumber
\end{eqnarray}
Here we can execute the trace in (\ref{fn1}) and we obtain
\begin{eqnarray}
&&\!\!\!\!\!\!\!\!\!\!\!\!\!\!\!\!\!\!\!\!\!
\Trni\eexp^{-\beta\hana{H}_n^{eff}(x^a)}
=\prod_a\int_0^\infty\dd x^a\eexp^{g^{RS}}\nonumber\\
&&\!\!\!\!\!\!\!\!\!\!\!\!\!\!\!\!\!\!
=\prod_a\int_0^\infty\dd x^a
   \exp\left\{
	B_1(\sum_ax^a)^2+B_2\sum_a(x^a)^2
        -B_3\sum_ax^a-B_4
   \right\}\nonumber\\
&&\!\!\!\!\!\!\!\!\!\!\!\!\!\!\!\!\!\!
=\prod_a\int_0^\infty\dd x^a
  \underbrace{
  \frac{1}{\sqrt{2\pi}}\int_{-\infty}^\infty\dd z \eexp^{-z^2/2}
             }_{\int\dd p_1(z)}
\exp\left\{
	\sqrt{2B_1}\sum_ax^az
	+B_2\sum_a(x^a)^2
        -B_3\sum_ax^a-B_4
       \right\}\nonumber\\
&&\!\!\!\!\!\!\!\!\!\!\!\!\!\!\!\!\!\!
=\int\dd p_1(z)\left[
  \int_0^\infty\dd x\exp\left\{
    zx\sqrt{2B_1}+B_2x^2-B_3x-\beta h\theta(y-x)
			\right\}
	       \right]^n_.\nonumber
\end{eqnarray}
At the final equality, we have replaced the $n$-fold multiple integral
of $x^a$ by a integral over a variable $x$ without the replica index $a$
because each integral of $x^a$ is independent each other.
Moreover, as we expect to take the limit $n\to 0$, using $a^n\simeq
1+n\ln a$A$\ln(1+n)\simeq n$, we obtain
\begin{eqnarray}
\ln\left[\Trni\eexp^{-\beta\hana{H}_n^{eff}(x^a)}\right]
&\simeq &\ln\int\dd p_1(z)\Bigg[
    1+n\ln\int_0^\infty\dd x\exp\Big\{
     B_2x^2 +(z\sqrt{2B_1}-B_3)x
  -\beta h\theta(y-x)
				\Big\}
		  \Bigg]\nonumber\\
&&\!\!\!\!\!\!\!\!\!\!\!\!\!\!\!\!\!\!\!\!\!
\!\!\!\!\!\!\!\!\!\!\!\!\!\!\!\!\!\!\!\!\!
=\ln\left[
  1+n\int\dd p_1(z)\ln\int_0^\infty\dd x\exp\{\cdots\}
    \right]
\nonumber
\simeq n\int\dd p_1(z)\ln\int_0^\infty\dd x\exp\{\cdots\}.
\end{eqnarray}
We, then, finally obtain the free energy density as
\begin{eqnarray}
f&=&\lim_{n\to 0}\frac{1}{n}\mbox{Min}f_n
=\lim_{n\to 0}\mbox{Min}\Bigg\{
    \frac{\beta \tilde{J}^2}{4}(n-1)q^2 + \frac{\beta \tilde{J}^2}{4}\sigma^2
    -\frac{Nm}{2}\tau^2 - \frac{r}{\beta}
    -\frac{1}{\beta}\int\dd p_1(z)\ln\int_0^\infty\dd x\eexp^{\beta g(x)}
                           \Bigg\}\nonumber\\
 &=&\mbox{Min}\Bigg\{
  -\frac{\beta \tilde{J}^2}{4}q^2 + \frac{\beta \tilde{J}^2}{4}\sigma^2 
  - \frac{Nm}{2}\tau^2 - \frac{r}{\beta}
 -\frac{1}{\beta}\int\dd p_1(z)\ln\int_0^\infty\dd x\eexp^{\beta g(x)}
              \Bigg\}_,\label{f}
\end{eqnarray}
where
\begin{equation}
g(x)\equiv 
   -\left(
    \frac{\beta \tilde{J}^2}{2}q-\frac{\beta \tilde{J}^2}{2}\sigma+\frac{u+m}{2}
   \right)x^2
+\left\{
    z\tilde{J}\sqrt{q}-(r/\beta+Nm\tau)
    \right\}x
   -h\theta(y-x).\nonumber
\end{equation}
Condition which minimizes $\tilde{f}$, which is the term in \{\} in the
right side of Eq. (\ref{f}) gives {\it mean field equations}.
As the term with $h$ is virtual for cumulative distribution function,
we let it to be zero hereafter. By the calculations,
$\partial\tilde{f}/\partial q=0, \quad
\partial\tilde{f}/\partial \sigma=0, \quad \partial\tilde{f}/\partial \tau=0,
\quad \partial\tilde{f}/\partial r=0$, we obtain the mean field equations,
\begin{eqnarray}
q&=&\int\dd p_1(z) \int_0^\infty\dd x
     \left(
      x^2 - \frac{zx}{\beta \tilde{J}\sqrt{q}}
     \right)
     \frac{\eexp^{\beta g(x)}}{Z_g}_,\label{q1}\\
\sigma&=&\int\dd p_1(z) \int_0^\infty\dd x
      \, x^2\, \frac{\eexp^{\beta g(x)}}{Z_g}_,\label{p1}\\
\tau&=&\int\dd p_1(z) \int_0^\infty\dd x
      \, x\, \frac{\eexp^{\beta g(x)}}{Z_g}=1_,\label{t1}
\end{eqnarray}
where $Z_g\equiv \int_0^\infty\dd x\exp (\beta g(x))$.
By substituting (\ref{p1}) for (\ref{q1}), we obtain
\begin{equation}
\sqrt{q}\tilde{J}\beta (\sigma-q)=\int\dd p_1(z)\, z\,
 \frac{\int_0^\infty\dd x\, x\exp(\beta g(x))}{Z_g}_.\label{q2}
\end{equation}
To estimate the integral over $x$ in the right side, let us
rewrite like following
\begin{eqnarray}
g(x)&=&-\underbrace{\frac{1}{2}\{(u+m)-\tilde{J}^2\beta (\sigma-q))\}}_{E_1}x^2
+ \underbrace{\{\tilde{J}\sqrt{q}z-(r/\beta + Nm \tau)\}}_{E_2}x\nonumber\\
    &=&-E_1 \left(x-\frac{E_2}{2E_1}\right)^2 + \frac{E_2^2}{4E_1}_.\label{gx}
\end{eqnarray}
The condition $E_1>0$ should be satisfied for the convergence of the
integral of $x$ in (\ref{q2}). When $E_2/(2E_1)>0$, that is, $z>(r/\beta
+Nm\tau)/(\tilde{J}\sqrt{q})\equiv -\Delta$ in (\ref{gx}), the integral of $x$ in 
Eq. (\ref{q2}) can be replaced by the integrand at the apex $x_1=E_2/2E_1$ in
the
limit $\beta\to\infty$. We, therefore, obtain
\begin{equation}
\int_0^\infty\dd x\, x\,\eexp^{\beta g(x)}=\left.\frac{E_2}{2E_1}\eexp^{\beta E_2^2/4E_1}\right._.\nonumber
\end{equation}
As 
\begin{equation}
Z_g=\int_0^\infty\dd x\,\eexp^{\beta g(x)}=\eexp^{\beta E_2^2/4E_1},\nonumber
\end{equation}
the contribution to (\ref{q2}) becomes $E_2/2E_1$.
Similarly, when $E_2/2E_1\leq 0$Athat is, $z\leq -\Delta$, the integral
of $x$ in Eq. (\ref{q2}) can be replaced by the integrand at
$x_2=0$, which is zero and has no contribution.
After the similar calculation of the integral in Eq. (\ref{t1}) and some transformation of the expressions, we obtain the following
mean field equations as
\begin{eqnarray}
 & & p-v = \sqrt{q}\int_{-\Delta}^\infty\dd p_1(z)(z+\Delta)_,
 \label{mfeq1}\nonumber\\
 & & (p-v)^2 = \int_{-\Delta}^\infty\dd p_1(z)(z+\Delta)^2_,
 \label{mfeq2}\nonumber\\
 & & \Delta = \sqrt{q}(p-2v)_,
  \label{mfeq3}\nonumber
\end{eqnarray}
where $v\equiv \tilde{J}\beta(\sigma-q)$. As stated previously, the above result is
essentially equivalent to the case $m=0, \tilde{J}=1$
\cite{Diederich_Opper_1989}. As the mean field equations can be solved
analytically only for some special values of $p$ (e.g., $p=\sqrt{2}$), we solve them
numerically for a general range of $p$ and obtain the order parameters $q, v$ as
a function of $p$, which are depicted in Figs. \ref{Fig1}(a) and 1(b). Macroscopic functions,
such as diversity and abundance distributions, are simultaneously obtained by substituting 
$q$ and $v$ for them.

%
%
\section{Diversity and abundance distributions}
Among macroscopic functions calculated in the present framework, the most significant for a
theory of RSA patterns is the cumulative distribution function of abundance (\ref{Cy})
which is derived from the free energy $f$ as 
\begin{eqnarray}
C_p (y)&\equiv&
  \lim_{h\to 0}\frac{\partial f}{\partial h}
=\lim_{h\to 0}\int\dd p_1(z)\int_0^\infty\dd x\,
 \theta(y-x)\eexp^{\beta g(x)}/Z_g\nonumber\\
&=&
   \int_{-\Delta}^\infty\dd p_1(z)\theta(y-E_2/(2E_1))
   +\int_{-\infty}^{-\Delta}\dd p_1(z) \theta(y),\nonumber
\end{eqnarray}
where the saddle point method was used in the last line like
in the integral over $x$ in (\ref{q2}). Substituting $E_1$ and $E_2$
for the above and rewriting the population by $x$ from $y$,
the resulting cumulative distribution function is represented as
\begin{equation}
C_p(x)=C_p(0)\theta(x)
    +\int_{-\Delta}^\infty\dd p_1(z)\theta\left(
	    x-\frac{\sqrt{q}(z+\Delta)}{p -v}
					  \right)_.\nonumber
\end{equation}
The quantity $C_p(0)\equiv\int_{-\infty}^{-\Delta}\dd p_1(z)$
gives cumulative distribution function of species with zero population,
that is, a proportion of extinct species. 

The function $\alpha_p(0)\equiv 1-C_{p}(0)=v(p-v)$ and $\alpha_p(1)=1-C_{p}(1)$ of
$p$ can be termed 'diversity', i.e., the proportion of nonextinct
species and that of the species with abundance larger than unity,
respectively, as depicted in Fig.~\ref{Fig2}. This demonstrates a
typical positive correlation between productivity and diversity
\cite{Waide_etal_1999}. Numerical results for $\alpha_p(1)$ are also
depicted in Fig.~\ref{Fig2} for comparison.  We see good agreement
between the analytical and the numerical results for $p\gtrsim 1$, while
some deviations appear for small values of $p$. This small-value
deviation is attributable to the occurrence of replica symmetry breaking
(RSB)\cite{Mezard_etal_1987,Biscari_Parisi_1995} for $p<\sqrt{2}$, which
yields a number of metastable states of Eq.~(\ref{f}), and the
replicator dynamics (\ref{eqReplicator}) essentially converges to not only a
ground state of (\ref{f}) but also to the metastable states. Since the
energy $\hana{H}$ and the diversity are both nonincreasing functions of
time in dynamics (\ref{eqReplicator}), the mean-field results here give a lower
minimum of diversity. Interestingly, the metastable states enhance the
diversity. The analysis of RSB is expected to improve the quantitative
agreement \cite{Biscari_Parisi_1995}.

In Fig.\ref{Fig2}, we also see a power law of the diversity $S\equiv
N\alpha_p(0)\propto p^\eta\, (p\lesssim 1;\, \eta\simeq 2.3)$. This can
be related to the species-area relationships
 $S\propto A^\lambda\,
(\lambda=\mbox{const.})$ \cite{MacArthur_Wilson_1967,May_1975} if $p$ is
a power function of area $A$, that is, larger the area, more producers
are observed than consumers, which is one of the predictions in the
present study. 

The function $\alpha_p (x)\equiv 1-C_p (x)$ is the survival function,
the proportion of species whose abundance is larger than $x$.
The
survival function has been often used in the medical statistics.
Note that $\alpha_p(x)$ is also represented as a
function of species rank $n$: 
\[
\alpha_p(x)=\frac{n}{N}\quad\mbox{for}\quad x\in [x^{(n+1)}, x^{(n)})_,
\]
$(n=1,2,\ldots , S\le N)$ if the species abundance is ranked in
descending order, as in $x^{(1)}\ge x^{(2)}\ge\cdots\ge
x^{(n)}\ge\cdots\ge x^{(S)}>0$. As the function $\alpha_p(x)$ is a
nonincreasing monotonic function, the species abundance relation, i.e., the abundance $x^{(n)}$ as a
function of a rank $n$, is given by the inverse function of
$\alpha_p(x)$ as 
$x^{(n)}=x_p(n/N)=\alpha_p^{(-1)}(x)$,
depicted in Fig.~\ref{Fig3} and Fig.~\ref{Fig4}
for some values of $p$. 

We observe two typical RSA patterns in different regions
\cite{Hubbell_2001} and with different species compositions
\cite{Whittaker_1970}: one is a straight line like the geometric series
\cite{Motomura_1932} for a small value of $p$, and the other consists of
sigmoid curves on a logarithmic vertical axis for some range of
$p$. This latter RSA pattern denotes a lognormal-like abundance
distribution. Remarkably, the transition of the RSA patterns from low $p$ to
high is identical to the observed transition from low- to
high-productivity areas; that is, from a species-poor area such as an
alpine or polar region to a species-rich tropical rain forest
\cite{Hubbell_2001}. The transition also corresponds to the secular
variation of RSA patterns observed in abandoned cultivated land
\cite{Bazzaz_1975}. This supports the contention that $p\; (\mbox{or} \, m)$ is a maturity parameter,
as is suggested by an evolutionary model \cite{Tokita_Yasutomi_2003}.

The abundance distribution is also derived from
the cumulative distribution function $C_p(x)$ by
its derivative as
\begin{eqnarray} 
F_p(x)&\equiv&\frac{\dd C_p(x)}{\dd x}\nonumber\\
&=&\frac{p-v}{\sqrt{2\pi q}}\exp\left\{
    -\frac{(p-v)^2}{2q}\left(
      x - \frac{q(p-2v)}{p-v}
                       \right)^2
                                \right\}
+C_p(0)\delta(x),\label{Fx}
\end{eqnarray}
and it is depicted 
in Fig.~\ref{Fig5} for some values of
$p$.
%
The first term is a normal distribution but not a lognormal
distribution. Nevertheless, the curves in Fig.~\ref{Fig4} demonstrate
a typical sigmoid pattern on a logarithmic vertical axis. This pattern
indicates the coexistence of very abundant species with rare ones. This
multiscale of abundance is intuitively understood by a divergent
behavior of the variance $\sigma^2\equiv q/(p-v)^2$ of $F_p(x)$ for
small $p$ because $q\to\infty$ and $v\to 0$ for $p\to 0$. Moreover, the
mode of $F_p(x)$ per 'natural' octave \cite{Preston_1962b} $\ln (x)$
is always a positive value (as shown in Fig.~\ref{Fig5}) at
$x^*=\frac{\sigma}{2}(\Delta + \sqrt{\Delta^2+4})>0$, which denotes a
unimodal distribution. Indeed, the mode diverges as
\[
x^*\to\frac{\sigma}{|\Delta |}=\frac{1}{(p-v)|p-2v|}\to\infty
\]
for $p\to 0$. As a result, the abundance
distribution is a normal distribution truncated at $x=0$ and given in the positive abundance range $x\ge 0$ and it
has a large variance $\sigma^2\to\infty$
and a negatively divergent mean $\mu\equiv\frac{q(p-2v)}{p-v}\to -\infty$
satisfying $\frac{\sigma}{\mu}=\frac{1}{\Delta}\to 0$ for $p\to 0$. 
This indicates that, for small $p$, $F_{p}(x)$ becomes a tail of broad distribution but it still has a peak at a positive
abundance $x^{*}$ when plotted on the log-scale horizontal axis.
This is why the abundance distribution per octave looks like a left-skewed lognormal
distribution \cite{Nee_Harvey_May_1991} in Fig.~\ref{Fig5}.

%
%
\section{Canonical hypothesis}

According to the {\it canonical hypothesis} \cite{Preston_1962b,May_1975}, the quantity
$\gamma\equiv\log(x_N)/\log(x_{max})$ takes a value near unity in various real communities,
where $x_{max}$ is the abundance of the most abundant species and $x_N$ gives the position of the mode of individual
curve $P_{p}(x)\equiv xF_{p}(x)$.
Using the abundance distribution (\ref{Fx}), we derive an analytical expression for the above functions
to check the validity of the canonical hypothesis. We first
evaluate an expected value of the most abundant species
$x_{max}$. From the definition of $x_{max}$, that is
$NF_p(x_{max})=1$, and the conservation of the total abundance
$\int_0^\infty xF_p(x)\dd x=N$, which is equivalent to
$\sum_i^Nx_i=N$, we obtain
\[
x_{max}=\frac{q(p-v)+\sigma\sqrt{2\ln\left(\frac{\sigma(1-\alpha_p(0))+\Delta\alpha_p(0)}{\sqrt{2\pi}}\right)}}{p-v}_.
\]
On the other hand, the mode of the individual curve $P_p(x)$ per
octave is given by $x_N=\frac{\sigma}{2}\left(\Delta
+\sqrt{\Delta^2+8}\right)$, and finally, the parameter
$\gamma$ is evaluated by substituting the
values of the order parameters $q$ and $v$ for each value of $p$.  In
the present model, $\gamma$ is a monotonically increasing function of
$p$ and $0.96<\gamma <1.04$ for $0.1<p<0.6$, denoting that the
canonical hypothesis is supported in the range of $p$ giving the
typical RSA patterns in Fig.~\ref{Fig5}. Although the canonical hypothesis
was demonstrated to be merely a mathematical consequence of lognormal
distribution \cite{May_1975} rather than anything biological, it is
noteworthy that the lognormal-like abundance distribution with
$\gamma\simeq 1$ derives from basic ecological dynamics. This still
suggests a biological foundation for the hypothesis in a large complex
ecosystem, in the same way that a biological foundation was indicated
for the theory of a local competitive community \cite{Sugihara_1980}.

%
%
\section{Topology of interactions}

The present theory seeks to capture the influence of productivity on the
RSA patterns under the assumption that all species interact randomly;
nevertheless, this assumption itself is never justified because it
ignores a biological correlation between interactions produced by
evolution. However, note that the randomness is assumed only for an
initial state with $N$ species in Eq.~(\ref{eqReplicator}). Actually, the
simulation reveals the resulting interactions of nonextinct species
to be nonrandom.

In Fig.~\ref{Fig6}, interspecies interactions ($b_{ij}$) between the non-extinct species 
of the corresponding LV equations (\ref{eqLV}) are depicted, which is
obtained by numerical simulations of RE (\ref{eqReplicator}) and the transformation (\ref{ri}) and (\ref{bij}).  The numerical integration of RE was executed for initial diversity $N=2048$, a randomly generated interspecies interactions $(J_{ij})$, a random initial population ($x_{i}$), and for (a) $p=0.1$ and (b) $0.2$ by the fourth-order Runge-Kutta method. At the equilibrium of (\ref{eqReplicator}) for the parameters, the number of non-extinct RE species $N\alpha_{p}(x=1)$ was (a) $11$ and (b) $28$. In each figure, the non-extinct LV species ($y_{i}>0$) without the resource species ($y_{M}=1$) are depicted by a blue disk which is arranged clockwise in descending order of the intrinsic growth rate $r_{i}$ as $1>r_1\ge r_2\ge\cdots\ge r_{i}\ge\cdots\ge r_{L}>0$ (hence every non-extinct LV species is a producer) where the number of non-extinct LV species is (a) $L=10$ and (b) $27$.
The diameter of the disk is in proportion to $|\log(r_1)|/|\log(r_i)|$. Each type of interaction is represented by its color: green links denote a mutualistic interaction $(b_{ij},b_{ji})=(+,+)$, yellow, a competitive $(-,-)$ and blue, an exploitation of more productive $i$ on less productive $j(>i)$ $(+,-)$ for $i<j$. No exploitation of less productive $j$ on more productive $i(<j)$ $(-,+)$ is observed (it were drawn by a red link). The thickness of each link is proportional to the larger value of $|b_{ij}|$ and $|b_{ji}|$.

It should be noted that not only each sample (a) and (b) but also every sample
calculated for Fig.~\ref{Fig2} evolved to only flora, $\forall i\,\,
r_i>0$. In every sample, only green, yellow and blue links are observed but no
red; thus the stable community after extinction dynamics obtains a hierarchical 
structure. Moreover, it is observed that more productive species with larger $r_{i}$ 
tend to have competitive (yellow) relationships each other and less productive ones have
mutualistic ones (green), the quantitative estimations of which is now in progress.
It is suggested that this emergent hierarchy is connected to the 
stability of a large-scale plant community with complex interspecies interactions of not only 
competition but also mutualism and exploitation.

%
%
\section{Discussions}

In the present model, all species coexist only in the limit
$p\to\infty$, that is, in the trivial cases in which interspecies
interactions are negligible ($\tilde{J}\ll u$ ) or homogeneous
($\tilde{J}\to 0$), thereby giving $\alpha_\infty (x) = \theta(1-x)$,
$x^{(n)}=x_\infty (n/N)=1$ for all $n$ and $F_\infty(x)=\delta(x-1)$.
Such a region of too large productivity, therefore, never corresponds with a real community even if all 
species coexist.
The symmetric interactions ($J_{ij}=J_{ji}$) considered in the present study really 
 emerged in a evolutionary community-assembly model\cite{Tokita_Yasutomi_2003}. 
 In the simulation, though the introduced mutants had asymmetric interactions
with existing species, the system evolved to have symmetric interactions. 
The system also showed a typical RSA pattern and therefore it can be said that the present study is an analytical treatment for it.
The system moreover was resistant to exotic species, which is manifested in the functional form of $v$ in Fig. \ref{Fig1}(b). 
 As the order parameter $v$ corresponds to {\it susceptibility} to external noise in the context of statistical physics, Fig. \ref{Fig1}(b) suggests that an ecosystem with medium productivity ($p\simeq\sqrt{2}$) is more sensitive to external disturbance than ones with lower or higher productivity. 
In other words, in the range $p\lesssim\sqrt{2}$ where the typical RSA patterns are observed, the lower $p$, the RSA patterns are more robust to external noise such as environmental change, which is one of the predictions of the present study and can be verified by field studies.

Simplicity of the present model with only one parameter conduces to some predictions 
to be verified by experimental researches.
They are summarized as follows. 
\begin{description}

\item[(1)] A RSA pattern of a community with not only competition but also mutualism and exploitation
shows itself like a tail of broad normal distribution. 
Although such a truncated normal distribution with negatively large
mean and large variance has not been examined to fit field data, there is still plenty of room to
consider alternative distributions for a community with complex interactions, to which the other models of competition, such as niche apportion models\cite{MacArthur_1957,Tokeshi_1999} or the neutral model\cite{Hubbell_2001}, is not applied. 

\item[(2)] Diversity (number of non-extinct species) $S$ is a power function of the productivity $p$ 
which is proportional to the average growth rate in Eq.~(\ref{defp}). 
As suggested in the first paragraph of this section, the present theory is not valid for a too large value
of the productivity, and the power law here may be applied to the
 left half of the hamp-shaped relationships with a peak at intermediate productivity levels,
 which has been reported most widely in field and experimental 
 researches\cite{Rosenzweig_1995,Tilman_Pacala_1993}.

\item[(3)] Productivity (the average growth rate) $p$ is a power function of area $A$. If it is, the
present theory also predicts the species-area relationships. 
The dependence of $p$ on $A$  appears against the intuition that the average growth rate 
per species is constant even if we enlarge an area of observation.
It should be, however, noted that such a constant $p$ is justified only if the
distribution $F_{p}(x)$ is invariable under changes of area $A$.
Although the present analysis assumes infinite total population $N$, thereby infinite area $A$, 
the finite-area effect on $p$ and $F_{p}(x)$ will be an important
subject to be addressed in future studies.

\item[(4)] The transitions of the RSA patterns from species-poor to species-rich community or 
from immature to mature community attribute to the productivity or the average growth rate $p$.
Such relationships between the various RSA patterns and ecological parameters will be one of 
the focal points of the next generation of community ecology though some classical models have no 
parameter and often gives no explanation on variations of RSA patterns.

\item[(5)] The canonical hypothesis is supported. Moreover, the value $\gamma$ is increasing function
of $p$.
Compared to the distribution $F_{p}(x)$ itself, the statistical evaluation of which is often controversial\cite{Harte_2003}, 
quantities like $\gamma$ seem to be more tractable in quantitative study of field data and there
still is a room for consideration of such 
macroscopic quantities which may characterize a large and complex community.

\item[(6)] A stable and complex community has a hierarchical structure in which more productive 
species exploits less ones, more productive species compete each other, and less productive species 
have mutualistic relationships among themselves. 
Exploring such a hierarchy and the bias of the competition and the mutualism will be one of the clues
to clarify the unsolved problems on the complexity and the stability in community ecology.

\end{description}
Verification of the above predictions are in progress through collaborations with field ecologists.

In summary, it has been demonstrated that empirically supported patterns
are derived from a single parameter of general population dynamics.
This not only suggests the
importance of globally coupled biological interactions in a large
assemblage but also provides a unified viewpoint on mechanisms of
similar patterns observed in other biological networks with
complex interactions; for example, a lognormal abundance distribution
of a protein in
cells \cite{Blake_etal_2003,Kaneko_PRE_2003,Sato_etal_2003}, which is
revealed by gene expression networks.

\section*{Acknowledgments}
The author thanks T. Chawanya and H. Irie for fruitful
discussions, and R. Frankham, Y. Iwasa, E. Matsen, R. May, M. Nowak and
J. Plotkin for their helpful comments.  The present study has been carried out under the
stimulating atmosphere of Large-scale Computational Science Division
(Kikuchi lab), Cybermedia Center, Osaka University, and Program for
Evolutionary Dynamics (Nowak lab), Harvard University. This work was
supported by Grants-in-Aid from MEXT, Japan (No.14740232, 17017019 and 17540383).

\bibliographystyle{junsrt}

\begin{thebibliography}{10}

\bibitem{Martinez_1991}
N~Martinez.
\newblock Artifacts or attributes? effects of resolution on the little rock
  lake food web.
\newblock {\em Ecological Monographs}, Vol.~61, pp. 367--392, 1991.

\bibitem{LittleRockLake}
{\tt http://userwww.sfsu.edu/\~{}webhead/lrl.html}.

\bibitem{BCPathway}
{\tt http://kr.expasy.org/cgi-bin/show\_thumbnails.pl}.

\bibitem{Jeong_et_al_2001}
H.~Jeong, S.~Mason, A.-L. Barab\'{a}si, and Z.~N. Oltvai.
\newblock Lethality and centrality in protein networks.
\newblock {\em Nature}, Vol. 411, pp. 41--42, 2001.

\bibitem{Gallery}
{\tt http://www.nd.edu/\~{}networks/gallery.htm}.

\bibitem{Barabasi_Linked_2002}
A~L Barab{\'{a}}si.
\newblock {\em Linked}.
\newblock Perseus, 2002.

\bibitem{Pastor-Satorras_Vespignani_2001}
R~Pastor-Satorras and A~Vespignani.
\newblock Epidemic dynamics and endemic states in complex networks.
\newblock {\em Phys. Rev. Lett.}, Vol.~63, p. 066117, 2001.

\bibitem{May_1999}
R.~M. May.
\newblock Unanswered questions in ecology.
\newblock {\em Phil. Trans. R. Soc. London}, Vol. {B 264}, pp. 1951--1959,
  1999.

\bibitem{Motomura_1932}
I.~Motomura.
\newblock On the statistical treatment of communities.
\newblock {\em Zoological Magazine, Tokyo}, Vol.~44, pp. 379--383, 1932.

\bibitem{Corbet_Fisher_Williams_1943}
A.~S. Corbet, R.~A. Fisher, and C.~B. Williams.
\newblock The relation between the number of species and the number of
  individuals in a random sample of an animal population.
\newblock {\em J. Anim. Ecol.}, Vol.~12, pp. 42--58, 1943.

\bibitem{MacArthur_1957}
R.~H. MacArthur.
\newblock On the relative abundance of bird species.
\newblock {\em Proc. Nat. Acad. Sci. USA}, Vol.~43, pp. 293--295, 1957.

\bibitem{MacArthur_1960}
R.~H. MacArthur.
\newblock On the relative abundance of species.
\newblock {\em Am. Nat.}, Vol.~94, pp. 25--36, 1960.

\bibitem{Preston_1962a}
F.~W. Preston.
\newblock The canonical distribution of commonness and rarity: Part 1.
\newblock {\em Ecology}, Vol.~43, pp. 185--215, 1962.

\bibitem{Preston_1962b}
F.~W. Preston.
\newblock The canonical distribution of commonness and rarity: Part 2.
\newblock {\em Ecology}, Vol.~43, pp. 410--432, 1962.

\bibitem{Whittaker_1970}
R~H Whittaker.
\newblock {\em {C}ommunities and {E}cosystems}.
\newblock Macmillan, New York, 1970.

\bibitem{Bazzaz_1975}
F~A Bazzaz.
\newblock Plant species diversity in oldfield successional ecosystems in
  southern illinois.
\newblock {\em Ecology}, Vol.~56, pp. 485--488, 1975.

\bibitem{May_1975}
R.~M. May.
\newblock {\em Patterns of species abundance and diversity}, pp. 81--120.
\newblock Belknap, Cambridge, 1975.

\bibitem{Sugihara_1980}
G~Sugihara.
\newblock Minimal community structure: an explanation of species abundance
  pattern.
\newblock {\em Am. Nat.}, Vol. 116, pp. 770--787, 1980.

\bibitem{Nee_Harvey_May_1991}
S~Nee, P~H Harvey, and R~M May.
\newblock Lifting the veil on abundance patterns.
\newblock {\em Proc. R. Soc. Lond. B}, Vol. 243, pp. 161--163, 1991.

\bibitem{Tokeshi_1999}
M.~Tokeshi.
\newblock {\em {S}pecies {C}oexistence}.
\newblock Blackwell, 1999.

\bibitem{Hubbell_2001}
S.~P. Hubbell.
\newblock {\em {T}he {U}nified {N}eutral {T}heory of {B}iodiversity and
  {B}iogeography}.
\newblock Princeton University Press, Princeton, 2001.

\bibitem{Hall_etal_2002}
M~Hall, K~Christensen, S~A {di Collabiano}, and H~J Jensen.
\newblock Time-dependent extinction rate and species abundance in a
  tangled-nature model of biological evolution.
\newblock {\em Phys. Rev. E}, Vol.~66, p. 011904, 2002.

\bibitem{Harte_2003}
J~Harte.
\newblock Tail of death and resurrection.
\newblock {\em Nature}, Vol. 424, pp. 1006--1007, 2003.

\bibitem{McGill_2003}
B.~J. McGill.
\newblock A test of the unified neutral theory of biodiversity.
\newblock {\em Nature}, Vol. 422, pp. 881--885, 2003.

\bibitem{Volkov_etal_2003}
I.~Volkov, J.~R. Banavar, S.~P. Hubbel., and A.~Maritan.
\newblock Neutral theory and relative species abundance in ecology.
\newblock {\em Nature}, Vol. 424, pp. 1035--1037, 2003.

\bibitem{Pigolotti_etal_2004}
S~Pigolotti, A~Flammini, and A~Maritan.
\newblock Stochastic model for the species abundance problem in an ecological
  community.
\newblock {\em Phys. Rev. E}, Vol.~70, p. 011916, 2004.

\bibitem{Etienne_Olff_2004}
R~S Etienne and H~Olff.
\newblock A novel genealogical approach to neutral biodiversity theory.
\newblock {\em Ecol. Lett.}, Vol.~7, pp. 170--175, 2004.

\bibitem{Chave_2004}
J~Chave.
\newblock Neutral theory and community ecology.
\newblock {\em Ecol. Lett.}, Vol.~7, pp. 241--253, 2004.

\bibitem{Harte_etal_EcolMonog_2005}
J~Harte, E~Conlisk, A~Ostling, J~L Green, and A~B Smith.
\newblock A theory of spatial structure in ecological communities at multiple
  spatial scales.
\newblock {\em Ecological Monographs}, Vol.~75, pp. 179--197, 2005.

\bibitem{Tokita_2004}
K.~Tokita.
\newblock Species abundance patterns in complex evolutionary dynamics.
\newblock {\em Phys. Rev. Lett.}, Vol.~93, pp. 178102--1$\sim$4, 2004.

\bibitem{Hofbauer_Sigmund_1998}
J~Hofbauer and K~Sigmund.
\newblock {\em {E}volutionary {G}ames and {P}opulation {D}ynamics}.
\newblock Cambridge University Press, Cambridge, 1998.

\bibitem{MacArthur_Levins_1967}
R.~H. MacArthur and R.~Levins.
\newblock The limiting similarity, convergence, and divergence of coexisting
  species.
\newblock {\em Am. Nat.}, Vol. 101, pp. 377--385, 1967.

\bibitem{May_1972}
R.~M. May.
\newblock Will a large complex system be stable?
\newblock {\em Nature}, Vol. 238, pp. 413--414, 1972.

\bibitem{Tokita_Yasutomi_2003}
K.~Tokita and A.~Yasutomi.
\newblock Emergence of a complex and stable ecosystem in replicator equations
  with extinction and mutation.
\newblock {\em Theor. Pop. Biol.}, Vol.~63, pp. 131--146, 2003.

\bibitem{Gardner_Ashby_1970}
M.~R. Gardner and W.~R. Ashby.
\newblock Connectance of large dynamic (cybernetic) systems - critical values
  for stability.
\newblock {\em Nature}, Vol. 228, p. 784, 1970.

\bibitem{Tokita_Yasutomi_1999}
K.~Tokita and A.~Yasutomi.
\newblock Mass extinction in a dynamical system of evolution with variable
  dimension.
\newblock {\em Phys. Rev. E}, Vol.~60, pp. 842--847, 1999.

\bibitem{Mezard_etal_1987}
M.~Mezard, G.~Parisi, and A.~Virasoro.
\newblock {\em {S}pin {G}lass {T}heory and {B}eyond}.
\newblock World Scientific, Singapore, 1987.

\bibitem{Diederich_Opper_1989}
S.~Diederich and M.~Opper.
\newblock Replicators with random interactions: {A} solvable model.
\newblock {\em Phys. Rev. A}, Vol.~39, pp. 4333--4336, 1989.

\bibitem{Biscari_Parisi_1995}
P.~Biscari and G.~Parisi.
\newblock Replica symmetry breaking in the random replicant model.
\newblock {\em J. Phys. A: Math. Gen.}, Vol.~28, pp. 4697--4708, 1995.

\bibitem{de_Oliveira_Fontanari_2000}
V.~M. {de Oliveira} and J.~F. Fontanari.
\newblock Random replicators with high-order interactions.
\newblock {\em Phys. Rev. Lett.}, Vol.~85, pp. 4984--4987, 2000.

\bibitem{de_Oliveira_Fontanari_2001}
V.~M. {de Oliveira} and J.~F. Fontanari.
\newblock Extinctions in the random replicator model.
\newblock {\em Phys. Rev. E}, Vol.~64, p. 051911, 2001.

\bibitem{de_Oliveira_Fontanari_2002}
V.~M. {de Oliveira} and J.~F. Fontanari.
\newblock Complementarity and diversity in a soluble model ecosystem.
\newblock {\em Phys. Rev. Lett}, Vol.~89, p. 148101, 2002.

\bibitem{Nishimori_2001}
H~Nishimori.
\newblock {\em Statistical Physics of Spin Glasses and Information Processing :
  An Introduction}.
\newblock Oxford University Press, 2001.

\bibitem{Waide_etal_1999}
{R B Waide, et al}.
\newblock The relationship between productivity and species richness.
\newblock {\em Annu. Rev. Ecol. Syst.}, Vol.~30, pp. 257--300, 1999.

\bibitem{MacArthur_Wilson_1967}
R.~H. MacArthur and E.~O. Wilson.
\newblock {\em {I}sland {B}iogeography}.
\newblock Princeton University Press, Princeton, 1967.

\bibitem{Rosenzweig_1995}
M~L Rosenzweig.
\newblock {\em {S}pecies {D}iversity in {S}pace and {T}ime}.
\newblock Cambridge Univ. Press, Cambridge, 1995.

\bibitem{Tilman_Pacala_1993}
D~Tilman and S~Pacala.
\newblock {\em {S}pecies {D}iversity in {E}cological {C}ommunities:
  {H}istorical and {G}eographical {P}erspectives}, pp. 13--25.
\newblock Univ. Chicago Press, New York, 1993.

\bibitem{Blake_etal_2003}
W.~J. Blake, M.~K{\ae}rn, C.~R. Cantor, and J.~J. Collins.
\newblock Noise in eukaryotic gene expression.
\newblock {\em Nature}, Vol. 422, pp. 633--637, 2003.

\bibitem{Kaneko_PRE_2003}
K~Kaneko.
\newblock Recursiveness, switching, and fluctuations in a replicating catalytic
  networks.
\newblock {\em Phys. Rev. E}, Vol.~68, p. 031909, 2003.

\bibitem{Sato_etal_2003}
K.~Sato, Y.~Ito, T.~Yomo, and K.~Kaneko.
\newblock On the relation between fluctuation and response in biological
  systems.
\newblock {\em Proc. Natl. Acad. Sci. USA}, Vol. 100, pp. 14086--14090, 2003.

\end{thebibliography}

\newpage
\appendix
\section{Average fitness as a Lyapunov function}
Since the interaction matrix $\bm{J}$ is symmetric ($J_{ij}=J_{ji}$), the time derivative is written as
\[
\frac{\dd\bar{f}(t)}{\dd t}=\left(\frac{\dd}{\dd t}\right)(\bm{x}\cdot\bm{Jx})
         =\left(\frac{\dd\bm{x}}{\dd t}\right)\cdot\bm{Jx}+\bm{x}\cdot\bm{J}\left(\frac{\dd\bm{x}}{\dd t}\right)
         =2\left(\frac{\dd\bm{x}}{\dd t}\right)\cdot\bm{Jx}.
\]
It is, therefore, found that
\begin{eqnarray}
\frac{1}{2}\frac{\dd\bar{f}(t)}{\dd t}&=&\sum_{i}\frac{\dd x_{i}}{\dd t}(\bm{Jx})_{i}
                                                                    =\sum_{i}x_{i}[(\bm{Jx})_{i}-\bm{x}\cdot\bm{Jx}](\bm{Jx})_{i}\nonumber\\
&=&\sum_{i}x_{i}(\bm{Jx})_{i}^{2}-\left[\sum_{i}x_{i}(\bm{Jx})_{i}\right]^{2}\nonumber\\
&=&\sum_{i}x_{i}\left[(\bm{Jx})_{i}-\bm{x}\cdot\bm{Jx}\right]^{2}\ge 0_{},\nonumber
\end{eqnarray}
and the average fitness $\bar{f}$ is non-decreasing function of time, a Lyapunov function.

\newpage
\section*{Figure Captions}

\begin{description}
\item[Figure \ref{Fig1}]  Order parameters (a) $q$ and (b) $v$ as a function of the productivity parameter $p$.

\item[Figure \ref{Fig2}] Diversity $\alpha_p(x=0,1)$ as a function of $p$ of log-log
scales for $x=0$(red) and $x=1$(green). Black circles show numerical solutions of $\alpha_p(1)$
averaged over randomly generated $50$ samples of $(J_{ij})$ for Eq.~(\ref{eqReplicator}) with
$N=2048$ and $p=0.1, 0.2, 0.3, 0.4, 0.5, \sqrt{2}/2, 1, \sqrt{2}, 2,
3$. Numerical integration of RE was executed by the fourth-order Runge-Kutta method. 
Error bars indicate the maximum and minimum values found in the samples.

\item[Figure \ref{Fig3}] Rank-abundance relations as a function of productivity $p$ on normal-normal scales.

\item[Figure \ref{Fig4}] Rank-abundance relations as a function of productivity $p$ on
normal-log scales.


\item[Figure \ref{Fig5}] Abundance distribution per 'natural' octave $\ln(x)$. Functions
$F_p(x)x(\dd (\ln(x))=F_p(x)\dd x)$ for some values of $p$ are depicted,
whereas Preston originally defined octaves as logarithms to base 2
\cite{Preston_1962b}.

\item[Figure \ref{Fig6}] Interspecies interactions ($b_{ij}$) between the non-extinct species of the corresponding LV equations.
\end{description}

\newpage

\begin{center}

\begin{figure}[h]
   \includegraphics[width=\textwidth]{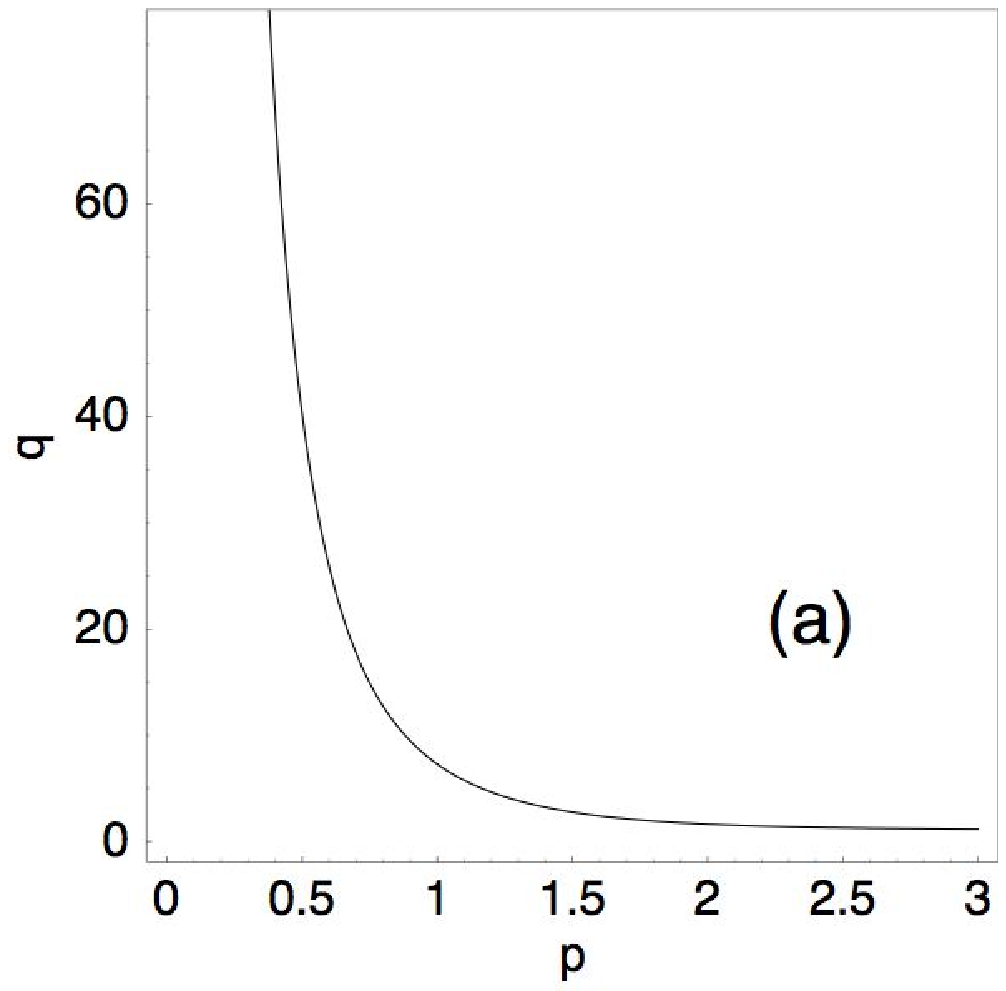}
   \caption{(a) TOKITA}
   \label{Fig1}
\end{figure}
\newpage
\setcounter{figure}{0}

\begin{figure}[h]
   \includegraphics[width=\textwidth]{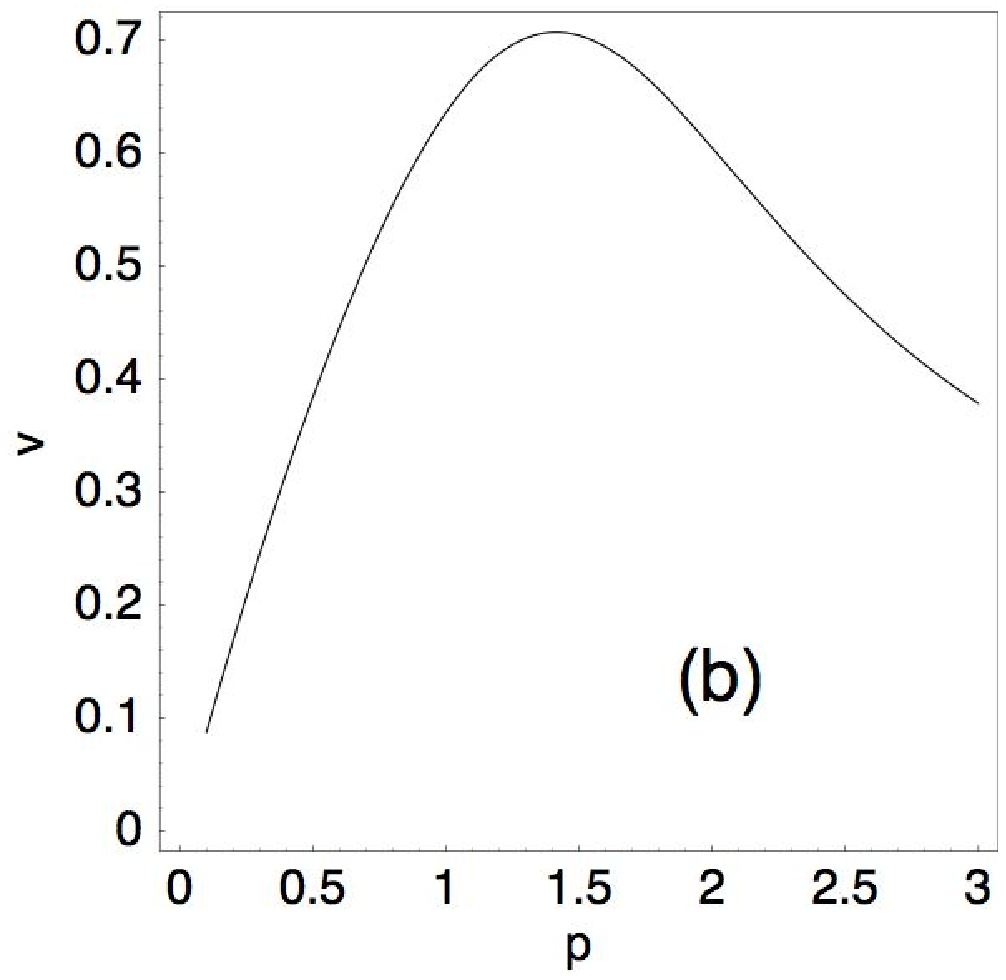}
   \caption{(b) TOKITA}
\end{figure}
\newpage

\begin{figure}[h]
   \includegraphics[width=\textwidth]{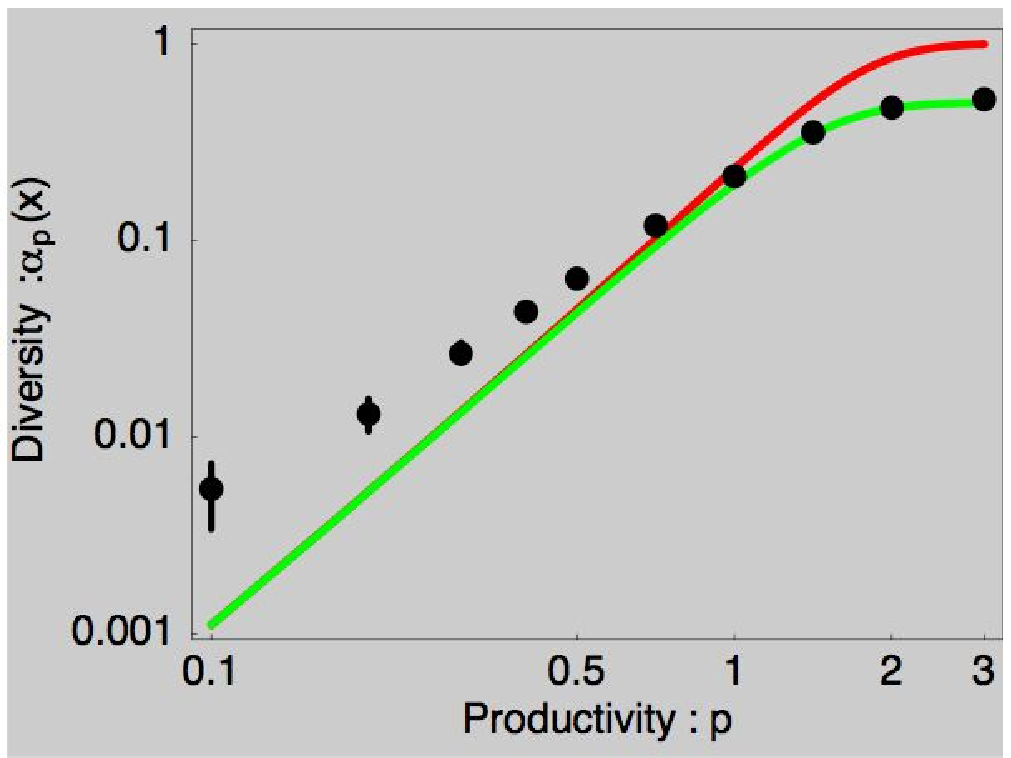}
   \caption{ TOKITA}
   \label{Fig2}
\end{figure}
\newpage

\begin{figure}[h]
   \includegraphics[width=\textwidth]{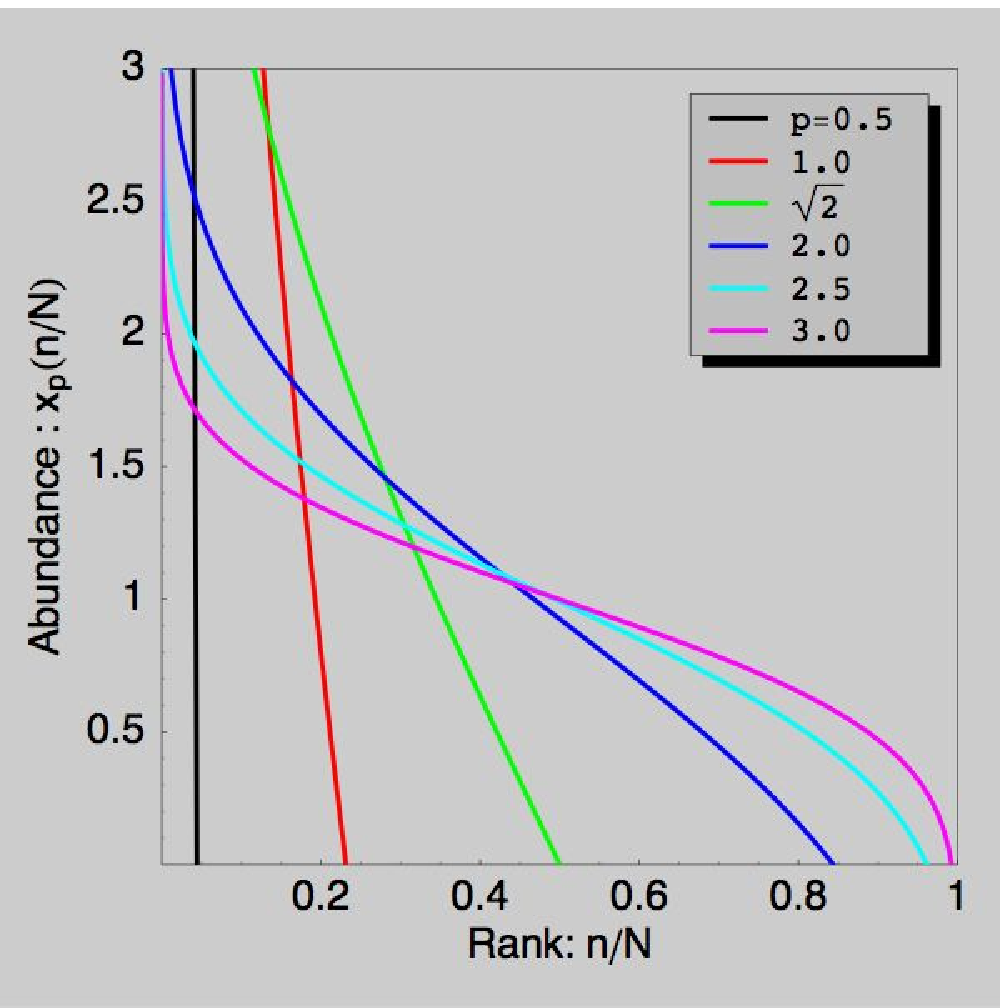}
   \caption{ TOKITA}
   \label{Fig3}
\end{figure}
\newpage

\begin{figure}[h]
   \includegraphics[width=\textwidth]{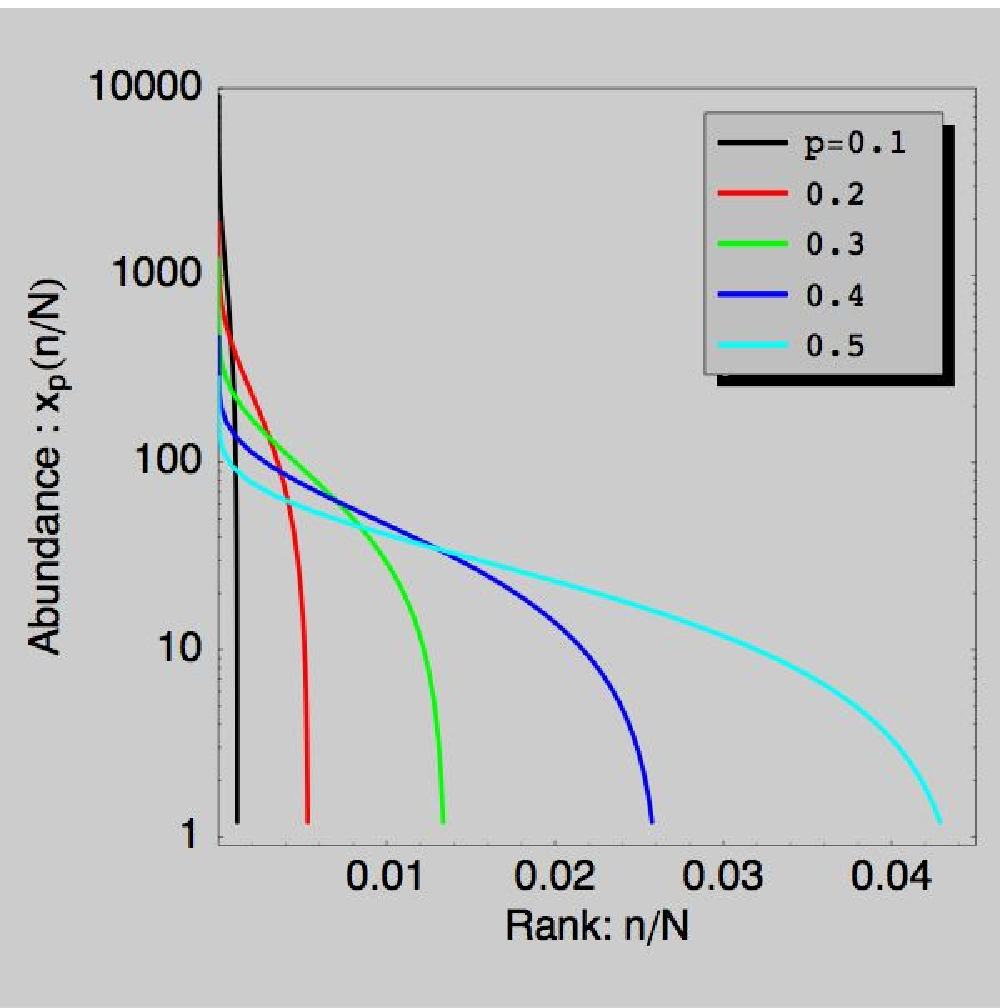}
   \caption{ TOKITA}
   \label{Fig4} 
\end{figure}
\newpage


\begin{figure}[h]
   \includegraphics[width=\textwidth]{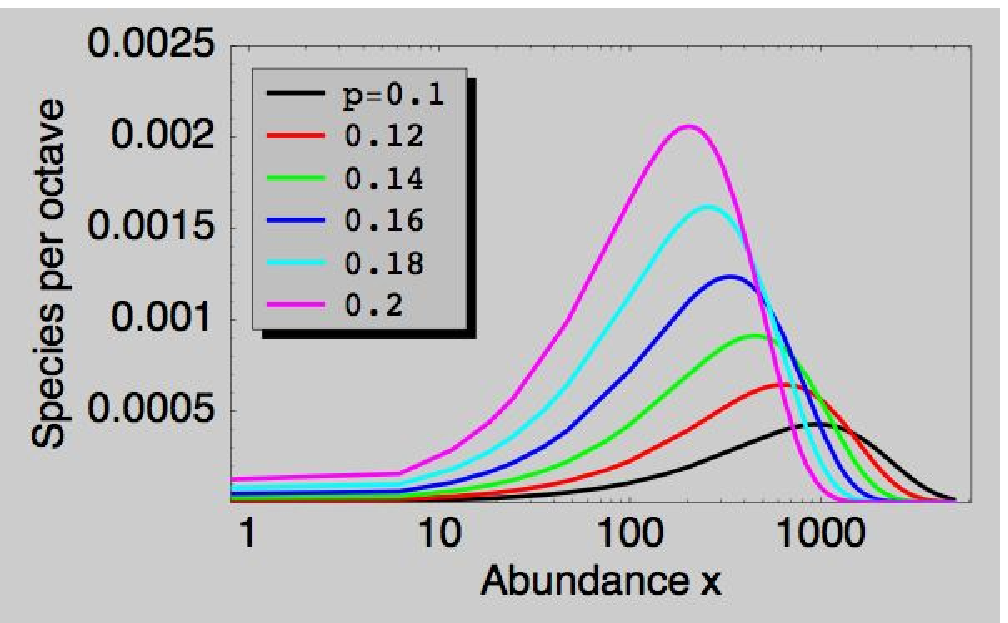}
   \caption{ TOKITA}
   \label{Fig5}
\end{figure}
\newpage

\begin{figure}[h]
   \includegraphics[width=\textwidth]{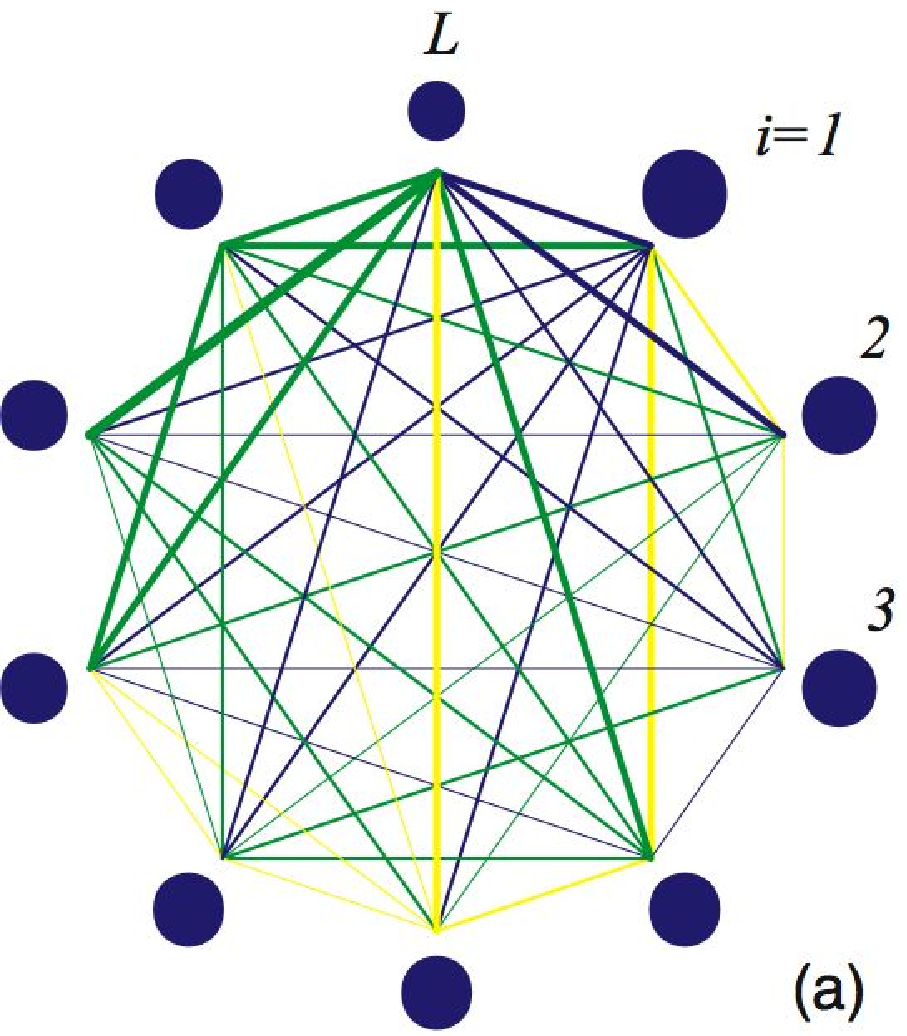}
   \caption{(a) TOKITA}
   \label{Fig6}
\end{figure}
\newpage

\setcounter{figure}{5}
\begin{figure}[h]
   \includegraphics[width=\textwidth]{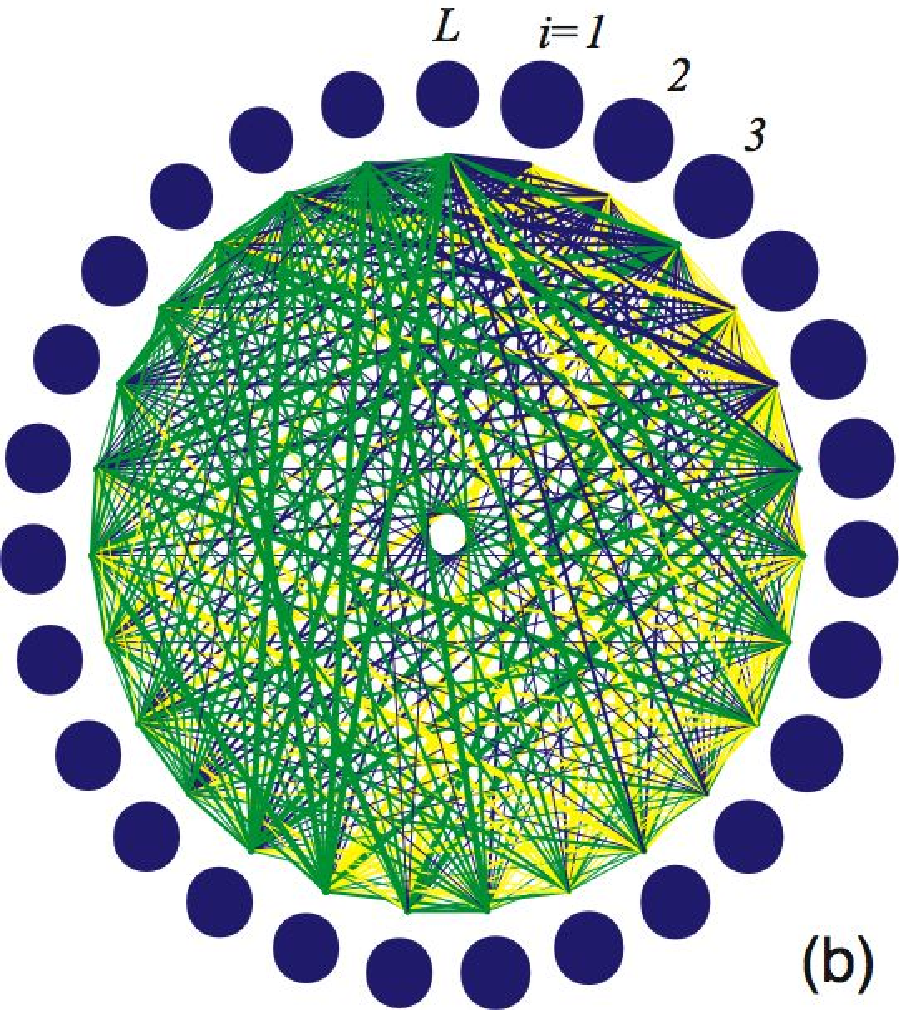}
   \caption{(b) TOKITA}
\end{figure}
\newpage

\end{center}

\end{document}